      [2004/11/16 v1.0c Il Nuovo Cimento]
\documentclass[a4paper]{cimento}

\usepackage[dvips]{graphicx}
\usepackage{geometry}
\usepackage[centertags]{amsmath}
\usepackage{amsfonts}
\usepackage{amssymb}
\usepackage{amsthm}
\usepackage{latexsym}
\begin{document}

\makeatletter
\def\@volume@{119 B}
\def\@issue@{1}
\def\@issuedate@{Gennaio 2004}
\makeatother

\setcounter{page}{9}

\checkin{il 26 Giugno 2002}{il 23 Febbraio 2004}

\title{Photon Number Emission in Synchrotron Radiation:
Systematics for High-Energy Particles}

\shorttitle{Photon Number Emission in Synchrotron Radiation: etc.}

\author{
    E.~B.~MANOUKIAN\from{ins:SUT}\thanks{Corresponding author, e-mail: edouard@ccs.sut.ac.th},
    N.~JEARNKULPRASERT\from{ins:SUT}\from{ins:TT} \atque
    P.~SUEBKA\from{ins:SUT}
}
\instlist{
  \inst{ins:SUT} School of Physics, Suranaree University of Technology,
                 Nakhon Ratchasima 30000, Thailand
  \inst{ins:TT} Astronomical Institute, Graduate School of Science,
                Tohoku Universtiy, Aoba, Aramaki, Sendai, 980-8578 Japan
}

\PACSes{
    \PACSit{41.60.-m}{Radiation by moving charges}
     \PACSit{41.60.Ap}{Synchrotron radiation}
     \PACSit{03.50.-z}{Classical fields theories}
     \PACSit{03.50.De}{Classical electromagnetism, Maxwell equations}
}

\maketitle

\begin{abstract}
A recent derivation of an explicit elementary expression for the
mean number $\left\langle N \right\rangle $ of photons emitted per
revolution in synchrotron radiation allows a sys\-te\-ma\-tic
high-energy analysis leading to the result
\begin{equation}
 \left\langle N \right\rangle\simeq
{{5\pi\alpha}/{\sqrt{3(1-\beta^2)}}}+a_0\alpha +
{{\pi\alpha\sqrt{1-\beta^2}}/{10\sqrt{3}}} \nonumber
\end{equation}
where $a_0$ is a constant, with relative errors of 2.2\%, .64\%,
.017\%, in com\-pa\-ri\-son to the well known formula tabulated in
the literature of 160\%, 82\%, 17\% for $\beta=0.8, 0.9, 0.99$,
respectively.
\end{abstract}

\section{Introduction.}
\noindent\indent Although many features of synchrotron radiation
have been well known for a long time (e.g., \cite{Pollock1983}), there is certainly
room for further improvements and developments. \ In a recent
investigation \cite{Manoukian2000}, starting from Schwinger's monumental work on
synchrotron radiation \cite{Schwinger1949} many years ago, an expicit expression
for the mean number $\left\langle N \right\rangle$ of photons
emitted per revolution was derived involving a remarkably simple
one-dimensional integral. \ This allowed us to carry out a
systematic analysis for high-energy charged particles. Our new
expression is given by
\begin{equation}
    \left\langle N \right\rangle   \simeq
    {{5\pi\alpha}/{\sqrt{3(1-\beta^2)}}}+ a_0\alpha +
    {{\pi\alpha\sqrt{1-\beta^2}}/{10\sqrt{3}}} \nonumber
\end{equation}
where $a_0$ is a numerical constant (see (\ref{Eq_13})). \ The
relative errors are $2.2\%$, $.64\%$, $.017\%$ in comparison to
the well known formula tabulated in the literature (see \cite{Particle1996})
involving only the first term in the above formula with relative
errors of $160\%$, $82\%$, $17\%$ for $\beta=0.8, 0.9, 0.99$,
respectively, which in turn has urged us and motivated us to
embark in the present investigation.

Our starting point for $\left\langle N \right\rangle  $ is derived
\cite{Manoukian2000} from the Schwinger expression for the power (\cite{Schwinger1949}
Eq.III.6, Eq.III.7)
giving:
\begin{equation}
    \left\langle N \right\rangle
    =\alpha\int^{\infty}_{0}\!\mathrm{d}z\int^{\infty}_{-\infty}\!\mathrm{d}x\;
    \mathrm{e}^{-\mathrm{i}zx}(\beta^2 \cos
    x-1)\frac{\sin(2z\beta\sin\frac{1}{2}x)}{\beta\sin\frac{1}{2}x}
    \label{Eq_01}
\end{equation}
Since the integrand factor in (\ref{Eq_01}), multiplying
$\exp(-\mathrm{i}zx)$, is an even function of $x$, only the real part of
the integral in (\ref{Eq_01}) contributes. \ It is easily verified
that $\left\langle N \right\rangle = 0$ for $\beta = 0$, as it
should be, when integrating over $x$ and $z$ in (\ref{Eq_01}) and by
using, in the process, that $\int^{\infty}_{0}\!\mathrm{d}z \
z\int^{\infty}_{-\infty}\!\mathrm{d}x\; \mathrm{e}^{-\mathrm{i}zx} = 0$. \
This latter boundary condition may be explicitly taken into
account for the vanishing of $\left\langle N \right\rangle  $ for
$\beta = 0$ by rewriting (\ref{Eq_01}) as
\begin{equation}
    \left\langle N \right\rangle
    =\alpha\int^{\infty}_{0}\!\mathrm{d}z\int^{\infty}_{-\infty}\!\mathrm{d}x\;
    \mathrm{e}^{-\mathrm{i}zx}\int^{\beta}_{0}\!\mathrm{d}\rho\left(\cos
    x\frac{\sin(2\beta z\sin\frac{1}{2}x)}{\sin\frac{1}{2}x} -
    \frac{2z}{\beta}\left[\cos\left(2z\rho\sin\frac{1}{2}x\right)-1\right]\right)\
    \label{Eq_02}
\end{equation}
We first integrate over $z$, then over $\rho$ and finally make a
change of variable $x/2\to z$ to obtain the remarkably
simple expression
\begin{equation}
    \left\langle N \right\rangle   =
    2\alpha\beta^2\int^{\infty}_{0}\!\frac{\mathrm{d}z}{z^2}\ \frac{\left(\frac{\sin
    z}{z}\right)^2 - \cos (2z)}{\left[1-\beta^2\left(\frac{\sin
    z}{z}\right)^2\right]}
    \equiv \alpha f(\beta)
    \label{Eq_03}
\end{equation}
There is no question of the existence of the latter integral for
all $0 \leqslant \beta <1$. \ The integral develops a
\emph{singularity} in $\beta$ for $\beta \to 1$.

\section{Systematic Treatment of $\left\langle N \right\rangle$ for High-Energy Particles.}
\noindent \indent Although the integral expression for
$\left\langle N \right\rangle  $ is simple, the investigation of
$\left\langle N \right\rangle  $ for $\beta \to 1$ is \emph{far}
\emph{from} trivial. \ For $\beta \to 1$, the integrand in
(\ref{Eq_01}) develops a \emph{singularity} for $z \to 0$. \ By
rewriting the integrand of $f(\beta)$ in (\ref{Eq_03}) in a form
suitable to study its behaviour for $z \to 0$ makes its
investigation for $z \to \infty$ rather difficult. \ Accordingly,
in Appendix A we have provided asymptotic expansions of some basic
functions involved in this work. \ Guided by these expansions, we
rewrite the integrand for $f(\beta)$ in (\ref{Eq_03}) explicitly as
\begin{align}
    \frac{2\beta^2}{z^2}\ \frac{\left(\frac{\sin z}{z}\right)^2 - \cos
    (2z)}{\left[1-\beta^2\left(\frac{\sin
    z}{z}\right)^2\right]} &= \frac{10 \beta}{3(1-\beta^2)+\beta^2
    z^2}+\frac{2}{z^2}\ \frac{6\left(\frac{\sin
    z}{z}\right)^2-\cos(2z)-5}{1-\left(\frac{\sin z}{z}\right)^2} \nonumber \\
    &\qquad -2(1-\beta^2)g(\beta,z)
    \label{Eq_04}
\end{align}
where
\begin{equation}
    g(\beta,z)=\frac{1}{z^2}\ \frac{\left[\left(\frac{\sin
    z}{z}\right)^2 - \cos (2z)\right]}{\left[1-\beta^2\left(\frac{\sin
    z}{z}\right)^2\right]\left[1-\left(\frac{\sin
    z}{z}\right)^2\right]} + \frac{5}{z^2}\ \frac{\left[\beta
    z^2-3(1+\beta)\right]}{3(1-\beta^2)(1+\beta)+\beta^2(1+\beta)z^2}
    \label{Eq_05}
\end{equation}
The essential point to note in (\ref{Eq_04}) is that we have factored
out $(1-\beta^2)$ in defining $g(\beta,z)$. \ The first term in
(\ref{Eq_04}) develops a singularity $1/z^2$ near the origin for
$\beta \to 1$. \ The second term in (\ref{Eq_04}) is independent of
$\beta$ and approaches a constant (see Eq.(\ref{Eq_A11}) for $z \to 0$,
and vanishes like $1/z^2$ for $z \to \infty$.

To investigate the contribution of $g(\beta,z)$ to $\left\langle N
\right\rangle  $, we rewrite the former as:
\begin{equation}
    g(\beta,z)=\frac{9}{30}\ \frac{\beta}{(1-\beta^2)+{\beta^2
    z^2}/{3}}-\frac{2}{3}\ \frac{(1-\beta^2)\beta}{\left[(1-\beta^2)+{\beta^2
    z^2}/{3}\right]^2}+g_1(\beta,z) \label{Eq_06}
\end{equation}
where
\begin{align}
    g_1(\beta,z) &= \frac{1}{z^2}\ \frac{\left[\left(\frac{\sin
    z}{z}\right)^2 - \cos (2z)\right]}{\left[1-\beta^2\left(\frac{\sin
    z}{z}\right)^2\right]\left[1-\left(\frac{\sin
    z}{z}\right)^2\right]} +
    \left[\frac{5}{3(1+\beta)}-\frac{9}{30}+\frac{2}{3}\
    \frac{(1-\beta^2)}{1-\beta^2+{\beta^2
    z^2}/{3}}\right]\nonumber \\
    &\qquad \times \frac{\beta}{\left[1-\beta^2+{\beta^2
    z^2}/{3}\right]}-\frac{5}{z^2}\ \frac{1}{1-\beta^2+{\beta^2
    z^2}/{3}}. \label{Eq_07}
\end{align}
In particular, we note that
\begin{equation}
    g_1(1,z)=\frac{1}{z^2}\left(\frac{\left(\frac{\sin z}{z}\right)^2
    - \cos (2z)}{\left[1-\left(\frac{\sin
    z}{z}\right)^2\right]^2}+\frac{8}{5}-\frac{15}{z^2}\right),
    \label{Eq_08}
\end{equation}
and that (see Eq.(\ref{Eq_A06}))
%\begin{equation}
%    g_1(1,z)\;{}_{\widetilde{z\to 0}}\;{-\frac{101}{600}} \label{Eq_09}
%\end{equation}
\begin{equation}
    g_1(1,z)\xrightarrow[\quad{}{z\to 0}\quad{}]{}{-\frac{101}{600}} \label{Eq_09}
\end{equation}
\begin{equation}
    g_1(1,z)\xrightarrow[\quad{}{z\to \infty}\quad{}]{}\mathcal{O}({1}/{z^2}).\label{Eq_10}
\end{equation}

Using the integrals:
\begin{equation}
    \int^{\infty}_{0}\!\frac{\mathrm{d}z}{[3+z^2]}=\frac{\pi}{2\sqrt{3}}
    \label{Eq_11}
\end{equation}
\begin{equation}
    \int^{\infty}_{0}\!\frac{\mathrm{d}z}{[3+z^2]^2}=\frac{\pi}{12\sqrt{3}}
    \label{Eq_12}
\end{equation}
\begin{equation}
    2\int^{\infty}_{0}\!\frac{\mathrm{d}z}{z^2}\ \frac{6\left(\frac{\sin
    z}{z}\right)^2-\cos(2z)-5}{\left[1-\left(\frac{\sin
    z}{z}\right)^2\right]}\equiv a_0 = -9.55797
    \label{Eq_13}
\end{equation}
we obtain from (\ref{Eq_03})--(\ref{Eq_13}) the explicit expression
\begin{equation}
    f(\beta)=\frac{5\pi}{\sqrt{3(1-\beta^2)}}+a_0+\frac{\pi}{10\sqrt{3}}\sqrt{1-\beta^2}+\varepsilon(\beta)
    \label{Eq_14}
\end{equation}
where
\begin{equation}
    \varepsilon(\beta) =
    -2(1-\beta^2)\int^{\infty}_{0}\!\mathrm{d}z \ g_1(\beta,z)
    \label{Eq_15}
\end{equation}
$a_0$ is given in (\ref{Eq_13}) and $g_1(\beta,z)$ is defined through
(\ref{Eq_07}), (\ref{Eq_08}) and (\ref{Eq_10}).

The error term $\varepsilon(\beta)$ is studied in Appendix B
giving
\begin{equation}
    \varepsilon(\beta) = \mathcal{O}[(1-\beta^2)] \label{Eq_16}
\end{equation}
for $\beta \to 1$.

Eq.(\ref{Eq_14}) may be also rewritten as
\begin{equation}
    f(\beta)=\frac{5\pi}{\sqrt{3(1-\beta^2)}}+a_0+\varepsilon_0(\beta)
    \label{Eq_17}
\end{equation}
where now the error $\varepsilon_0(\beta)$ is from (\ref{Eq_14}) and
(\ref{Eq_16})
\begin{equation}
    \varepsilon_0(\beta)=\mathcal{O}(\sqrt{1-\beta^2}) \label{Eq_18}
\end{equation}
for $\beta \to 1$.

According to Eqs.(\ref{Eq_14}) and (\ref{Eq_17}), we obtain the
following representation for $\left\langle N \right\rangle  $ for
high-energy charged particles:
\begin{equation}
    \left\langle N \right\rangle   \simeq \frac{5\pi\alpha}{\sqrt{3(1-\beta^2)}}+a_0\alpha
    +\frac{\pi\alpha\sqrt{1-\beta^2}}{10\sqrt{3}} \label{Eq_19}
\end{equation}
where the numerical constant $a_0$ is given in (\ref{Eq_13}), with
relative errors as mentioned in the the Introduction to be
compared with the well known tabulated formula \cite{Particle1996}. \ It is
important to emphasize that the asymptotic constant $a_0$ is
overwhelmingly large in magnitude and may be easily missed in a
non-systematic analysis and is the very important contribution in
(\ref{Eq_19}).

\section*{Appendix A}
\renewcommand{\theequation}{A.\arabic{equation}}
\setcounter{equation}{0}

\noindent\indent We provide asymptotic expansions of some basic
functions involved in this work for $z \to 0$. \ To this end we
note the following expansions:
\begin{equation}
    \left(\frac{\sin
    z}{z}\right)^2-\cos(2z)\cong\frac{5}{3}z^2-\frac{28}{45}z^4+\frac{31}{360}z^6+\cdots
    \label{Eq_A01}
\end{equation}
\begin{equation}
     \left(\frac{\sin
    z}{z}\right)^2\cong 1-\frac{z^2}{3}+\frac{2}{45}z^4-\frac{z^6}{360}+\cdots
    \label{Eq_A03}
\end{equation}
\begin{equation}
    \left[1-\left(\frac{\sin
    z}{z}\right)^2\right]^2\cong\frac{z^4}{9}\left(1-\frac{4}{15}z^2+\frac{31}{900}z^4\right)+\cdots
    \label{Eq_A03}
\end{equation}
These expressions lead to the following asymptotic expansions for
the functions:
\begin{equation}
    f_1(z)=\frac{\left(\frac{\sin z}{z}\right)^2
    - \cos (2z)}{\left[1-\left(\frac{\sin z}{z}\right)^2\right]^2}
    \label{Eq_A04}
\end{equation}
\begin{equation}
    f_2(z)=\frac{\left(\frac{\sin z}{z}\right)^2
    - \cos (2z)}{\left[1-\left(\frac{\sin
    z}{z}\right)^2\right]^2}\left(\frac{\sin z}{z}\right)^2,
    \label{Eq_A05}
\end{equation}
\begin{equation}
    f_1(z)\xrightarrow[\quad{}{z\to 0}\quad{}]{}\frac{15}{z^2}-\frac{8}{5}
    -\frac{101}{600}z^2+\mathcal{O}(z^4)
    \label{Eq_A06}
\end{equation}
\begin{equation}
    f_2(z)\xrightarrow[\quad{}{z\to 0}\quad{}]{}\frac{15}{z^2}-\frac{33}{5}
    +\frac{619}{600}z^2+\mathcal{O}(z^4)
    \label{Eq_A07}
\end{equation}
and the following asymptotic expansions for the functions:
\begin{equation}
    f_3(z)=\frac{6\left(\frac{\sin z}{z}\right)^2
    - \cos (2z)-5}{1-\left(\frac{\sin z}{z}\right)^2}
    \label{Eq_A08}
\end{equation}
\begin{equation}
    f_4(z)=\frac{\left(\frac{\sin z}{z}\right)^2
    - \cos (2z)}{\left[1-\left(\frac{\sin
    z}{z}\right)^2\right]^2}\left(\frac{\sin z}{z}\right)^2
    \left[\frac{z^2}{3}+\left(\frac{\sin z}{z}\right)^2-1\right]
    \label{Eq_A09}
\end{equation}
\begin{equation}
    f_5(z)\equiv\frac{\left(\frac{\sin z}{z}\right)^2
    - \cos (2z)}{\left[1-\left(\frac{\sin
    z}{z}\right)^2\right]^2}\left(\frac{\sin z}{z}\right)^2\left[\frac{z^2}{3}+\left(\frac{\sin
    z}{z}\right)^2-1\right]^2,
    \label{Eq_A10}
\end{equation}
\begin{equation}
    f_3(z)\xrightarrow[\quad{}{z\to 0}\quad{}]{}-\frac{6}{5}z^2+\mathcal{O}(z^4)
    \label{Eq_A11}
\end{equation}
\begin{equation}
    f_4(z)\xrightarrow[\quad{}{z\to 0}\quad{}]{}\frac{2}{3}z^2-\frac{67}{200}z^4
    +\mathcal{O}(z^6)
    \label{Eq_A12}
\end{equation}
\begin{equation}
    f_5(z)\xrightarrow[\quad{}{z\to 0}\quad{}]{}\frac{4}{135}z^6+\mathcal{O}(z^8)
    \label{Eq_A13}
\end{equation}

\section*{Appendix B}
\renewcommand{\theequation}{B.\arabic{equation}}
\setcounter{equation}{0}

\noindent\indent In this appendix we investigate the nature of the
error term in (\ref{Eq_14}). \ To this end, we rewrite the integrand
$g_1(\beta,z)$ defined in (\ref{Eq_07}) as
\begin{equation}
    g_1(\beta,z)=[g_1(\beta,z)-g_1(1,z)]+g_1(1,z),
    \label{Eq_B01}
\end{equation}
\begin{align}
    g_1(\beta,z)-g_1(1,z) &= -\frac{(1-\beta^2)}{z^2}\frac{f_2(z)}{\left[1-\beta^2\left(\frac{\sin
    z}{z}\right)^2\right]}+\frac{2}{3}\frac{(1-\beta^2)\beta}{[1-\beta^2+{\beta^2
    z^2}/{3}]^2} \nonumber \\
    &\qquad +\frac{57}{120}\frac{(1-\beta^2)\beta}{[1-\beta^2+{\beta^2
    z^2}/{3}]}+\frac{(1-\beta^2)^2\beta (107+57\beta)}{120(1+\beta)^3[1-\beta^2+{\beta^2
    z^2}/{3}]} \nonumber \\
    &\qquad -\frac{33}{5z^2}\frac{(1-\beta^2)}{[1-\beta^2+{\beta^2
    z^2}/{3}]} + \frac{15}{z^4}\frac{(1-\beta^2)}{[1-\beta^2+{\beta^2
    z^2}/{3}]}
    \label{Eq_B02}
\end{align}
and where $f_2(z)$ is defined in (\ref{Eq_A05}).

Upon using the identity
\begin{align}
    \frac{1}{\left[1-\beta^2\left(\frac{\sin
    z}{z}\right)^2\right]} &= \frac{\beta^6\left[\frac{z^2}{3}+\left(\frac{\sin
    z}{z}\right)^2-1\right]^3}{[1-\beta^2+{\beta^2
    z^2}/{3}]^3\left[1-\beta^2\left(\frac{\sin
    z}{z}\right)^2\right]}+ \frac{\beta^4\left[\frac{z^2}{3}+\left(\frac{\sin
    z}{z}\right)^2-1\right]^2}{[1-\beta^2+{\beta^2
    z^2}/{3}]^3} \nonumber \\
    &\qquad +\frac{\beta^2\left[\frac{z^2}{3}+\left(\frac{\sin
    z}{z}\right)^2-1\right]}{[1-\beta^2+{\beta^2
    z^2}/{3}]^2}+\frac{1}{[1-\beta^2+{\beta^2
    z^2}/{3}]}
    \label{Eq_B03}
\end{align}
we may rewrite (\ref{Eq_B02}) as
\begin{equation}
    g_1(\beta,z)-g_1(1,z)=\sum\limits^{5}_{i=1}F_i(\beta,z)
    \label{Eq_B04}
\end{equation}
where
\begin{equation}
    F_1(\beta,z)=-\frac{(1-\beta^2)\beta^6 f_5(z)\left[\frac{z^2}{3}+\left(\frac{\sin
    z}{z}\right)^2-1\right]}{z^2\left[1-\beta^2\left(\frac{\sin
    z}{z}\right)^2\right][1-\beta^2+{\beta^2
    z^2}/{3}]^3}
    \label{Eq_B05}
\end{equation}
\begin{equation}
    F_2(\beta,z)=-\frac{(1-\beta^2)\beta^4 f_5(z)}{z^2[1-\beta^2+{\beta^2
    z^2}/{3}]^3}
    \label{Eq_B06}
\end{equation}
\begin{equation}
    F_3(\beta,z)=-\frac{(1-\beta^2)\beta^4
    f_4(z)}{z^2[1-\beta^2+{\beta^2
    z^2}/{3}]^2}+\frac{2}{3}\frac{(1-\beta^2)}{[1-\beta^2+{\beta^2
    z^2}/{3}]^2}
    \label{Eq_B07}
\end{equation}
\begin{equation}
    F_4(\beta,z)=-\frac{(1-\beta^2)(f_2(z)+{{33}/{5}}-{{15}/{z^2}})}{z^2[1-\beta^2+{\beta^2
    z^2}/{3}]}
    \label{Eq_B08}
\end{equation}
\begin{equation}
    F_5(\beta,z)=\frac{57}{120}\frac{(1-\beta^2)\beta}{[1-\beta^2+{\beta^2
    z^2}/{3}]}+\frac{(1-\beta^2)^2\beta (107+57\beta)}{120(1+\beta)^3[1-\beta^2+{\beta^2
    z^2}/{3}]}
    \label{Eq_B09}
\end{equation}
and $f_2(z)$, $f_4(z)$, $f_5(z)$ are defined, respectively, in
(\ref{Eq_A05}), (\ref{Eq_A09}), (\ref{Eq_A10}). \ From the asymptotic expansions of the
latter functions in (\ref{Eq_A07}), (\ref{Eq_A12}), (\ref{Eq_A13}), respectively, we infer
that $g_1(\beta,z)-g_1(1,z)$ gives a
$\mathcal{O}\left((1-\beta^2)^{3/2}\right)$ contribution to
$\varepsilon(\beta)$. \ On the other hand, from
(\ref{Eq_08})--(\ref{Eq_10}), we infer that $g_1(1,z)$ in (\ref{Eq_B01}) will give
a $\mathcal{O}\left((1-\beta^2)\right)$ contribution to
$\varepsilon(\beta)$. \  All told we infer that
$\varepsilon(\beta)=\mathcal{O}\left((1-\beta^2)\right)$ for
$\beta \to 1$.

\end{document}